# Intrinsic origin of interfacial second-order magnetic anisotropy in ferromagnet/normal metal heterostructures

Hyung Keun Gweon[1], Hyeon-Jong Park[2], Kyoung-Whan Kim[3], Kyung-Jin Lee[1,2]* & Sang Ho Lim[1]†

[1]*Department of Materials Science and Engineering, Korea University, Seoul 02841, Korea*

[2]*KU-KIST Graduate School of Converging Science and Technology, Korea University, Seoul 02841, Korea*

[3]*Center for Spintronics, Korea Institute of Science and Technology, Seoul 02792, Korea*

*,†Corresponding authors. Email: †sangholim@korea.ac.kr, *kj_lee@korea.ac.kr

**Interfacial perpendicular magnetic anisotropy, which is characterized by the first-order ($K_1$) and second-order ($K_2$) anisotropies, is the core phenomenon for nonvolatile magnetic devices. A sizable $K_2$ satisfying a specific condition stabilizes the easy-cone state, where equilibrium magnetization forms at an angle from the film normal. The easy-cone state offers intriguing possibilities for advanced spintronic devices and unique spin textures, such as spin superfluids and easy-cone domain walls. Experimental realization of the easy-cone state requires understanding the origin of $K_2$, thereby enhancing $K_2$. However, previously proposed origins of $K_2$ cannot fully account for experimental results. Here we show experimentally that $K_2$ scales almost linearly with the work-function difference between the Co and X layers in Pt/Co/X heterostructures (X = Pd, Cu, Pt, Mo, Ru, W, and Ta), suggesting the central role of the inversion asymmetry in $K_2$. Our result provides a guideline for enhancing $K_2$ and realizing magnetic applications based on the easy-cone state.**

**Introduction**

Magnetic anisotropy describes a magnetization-angle-dependent change in magnetic energy and stabilizes the magnetization in specific directions. Its angular dependence is determined by the symmetry of the crystal or structure. In thin-film heterostructures such as ferromagnet/normal metal bilayers where the structural inversion symmetry is broken at the interface, the magnetic anisotropy is dominated by interfacial contributions, as follows (up to the second order):

$$E(\theta) = K_1^{\text{eff}} \sin^2 \theta + K_2 \sin^4 \theta. \tag{1}$$

Here, $K_1^{\text{eff}} (= K_1 - 2\pi M_s^2)$ is the effective first-order anisotropy energy density that comprises the demagnetization energy density (with $K_1$ and $M_s$ being the first-order anisotropy energy density and saturation magnetization, respectively), $K_2$ is the second-order anisotropy energy density, and $\theta$ is the polar angle of the magnetization. The magnetic phase diagram as functions of $K_1^{\text{eff}}$ and $K_2$ (Fig. 1a) shows several distinct magnetic states[1]. Among them, the out-of-plane state originating from perpendicular magnetic anisotropy (PMA) has been a main focus of spintronics research[2] because it offers scalable magnetic random-access memories (MRAMs)[3].

Recently, interest in another state—the easy-cone state, where the equilibrium magnetization direction is tilted from the film normal and forms a cone—has increased for the following reasons. It provides improved functionalities of various spintronics devices, such as low-power operation of spin-transfer torque (STT) MRAMs[4–6] and zero-field precession of STT oscillators[7]. Moreover, it hosts spin superfluids associated with spontaneous breaking of $U(1)$ spin-rotational symmetry[8,9] and allows unique easy-cone domain wall dynamics[10]. The existence of the easy-cone state was experimentally verified in various layered structures[6,11,12]. However, the design window for forming a stable easy-cone state is very narrow[6,11,12], which presents a critical challenge for realizing magnetic devices utilizing the easy-cone state.

In contrast to the out-of-plane state that can form with $K_1$ alone, the easy-cone state requires a large $K_2$ value; it is formed for $K_1^{\text{eff}} < 0$ and $K_2 > -1/2\, K_1^{\text{eff}}$ (Fig. 1a). In order to actively employ the easy-cone state in various applications, therefore, it is of crucial importance to find a way of enhancing $K_2$, which necessitates a fundamental understanding of its origin. The origin of $K_1$ has long been a subject of extensive theoretical and experimental research. It was found to depend on the orbital anisotropy[13], spin–orbit interaction of electronic structures near the Fermi level[14], or Rashba-type spin–orbit interaction at the interface associated with the inversion symmetry breaking[15–17]. Concerning the origin of $K_2$, three mechanisms have been proposed: 1) spatial fluctuations of $K_1$[18], 2) interfacial PMA combined with a gradual weakening of the exchange energy along the thickness direction[19], and 3) the mixture of bulk magnetocrystalline cubic anisotropy and interfacial uniaxial anisotropy[20]. The first and second mechanisms predict only positive $K_2$ and fail to explain the negative $K_2$ observed in experiments[21,22]. The third mechanism predicts both signs of $K_2$ depending on the nature of the *bulk* cubic anisotropy. Our measurement of $K_2$ for a Pt/Co/Cu structure, however, shows that $K_2$ is inversely proportional to the Co thickness (thus, the interface origin) and is negative for thin Co layers (see Fig. 1c and Supplementary Note 1 for details). As the third mechanism cannot account for the negative $K_2$ of interface origin, none of the three aforementioned mechanisms can explain this experimental observation; thus, a new origin of $K_2$ must be identified.

In this study, we focus on the role of inversion symmetry breaking in $K_2$, for the following two reasons. First, recent theoretical and experimental studies indicated an important role of the inversion asymmetry in $K_1$ for ferromagnet/normal metal heterostructures[15–17]. As $K_1$ and $K_2$ are the order-expanded coefficients of the net magnetic anisotropy [equation (1)], it is reasonable to expect that they share the same origin. Second, our measurements of $K_1$ and $K_2$ for Pt/Co/Cu and Pt/Co/MgO stacks over a wide range of Co thicknesses ($t_{Co}$) show that for both $K_1$ and $K_2$, the interfacial contribution is dominant compared with the bulk one (Supplementary Note 1), indicating the important role of the inversion asymmetry at the interface in the anisotropy.

**Materials and methods**

**Sample preparation**

To investigate the correlation between the inversion asymmetry and $K_2$, we examine various sputtered Pt/Co/X stacks, with X = Pd, Cu, Pt, Mo, Ru, W, and Ta. The stacks investigated in this study had the structure of Si substrate (wet-oxidized)/Ta (5 nm)/Pt (5 nm)/Co (1 nm)/X (3 nm)/Ta (3 nm) and were fabricated using an ultrahigh-vacuum magnetron sputtering system with a base pressure of $8 \times 10^{-8}$ Torr. All the metallic layers were deposited under an Ar pressure of $2 \times 10^{-3}$ Torr. The Ta under- and upper-layers were introduced to improve the surface roughness and prevent the oxidation of the stacks, respectively. For X = Ta, Pt (3 nm) was used as the upper-layer. Pt/Co/MgO (2 nm) stacks were also prepared, followed by post-annealing at 400°C for 30 min to maximize the interfacial PMA at the Co/MgO interface[23–25]. Details regarding the fabrication and annealing are provided in Supplementary Note 4. The continuous samples were patterned into a Hall bar structure via photolithography and inductively coupled plasma etching. The current-injection line and the voltage branch had dimensions of 5 μm (width) × 35 μm (length). A 50-nm-thick Pt layer was deposited on top of the patterned structure as a contact pad for magnetotransport characterization (Fig. 2a).

**Measurement of magnetic anisotropy**

The magnetic anisotropies ($K_1$ and $K_2$) were characterized by the anomalous Hall effect (AHE) in a standard four-probe Hall geometry. The Hall bar device was mounted on a rotatable sample stage placed in the gap of an electromagnet. The AHE measurements involved injecting an in-plane current ($I_x$ = 5 mA) along the $x$ direction and sensing the Hall voltage induced along the $y$ direction. The external magnetic field ($H_{ext}$) was applied at a polar angle ($\theta_H$) of 80° to facilitate coherent magnetization behaviour (Fig. 2a). The generalized Sucksmith–Thompson method was used to accurately determine the effective first- and second-order anisotropy fields (denoted as $H_{K1}^{eff}$ and $H_{K2}$, respectively)[26]. The key to this method is the use of the following equations, which can be derived from the total magnetic energy equation [equation (1), considering the Zeeman energy ($-\mathbf{M}\cdot\mathbf{H}_{ext}$)]:

$$\alpha H_{ext} = H_{K1}^{eff} + H_{K2}(1-m_z^2), \tag{2}$$

$$\alpha \equiv \frac{m_z \sin\theta_H - \sqrt{1-m_z^2}\cos\theta_H}{m_z\sqrt{1-m_z^2}}. \tag{3}$$

The AHE results were normalized with respect to the anomalous Hall voltages to obtain $m_z$–$H_{ext}$ curves (Fig. 2b), and then $\alpha H_{ext}$ was plotted with respect to $1-m_z^2$ to extract $H_{K1}^{eff}$ and $H_{K2}$ from the intercept and slope, respectively [equation (2) and Fig. 2c]. We observed a slight misalignment in $\theta_H$ from its nominal value (mostly within 2°), which was adjusted to maximize the linearity of the $\alpha H_{ext}$ vs. $1-m_z^2$ plot. To confirm the accuracy of the anisotropy constants, the measured $m_z$–$H_{ext}$ curves were compared with those from macrospin simulations using the obtained $H_{K1}^{eff}$ and $H_{K2}$ values as inputs (Fig. 2b). The $M_s$ values of the continuous samples were measured using a vibrating sample magnetometer. The anisotropy constants were then obtained from the relationships $K_1 = M_s H_{K1}^{eff}/2 + 2\pi M_s^2$ and $K_2 = M_s H_{K2}/4$. All the measurements were performed at room temperature.

**Measurement of work function**

To measure the work functions of the metals and MgO, ultraviolet photoelectron spectroscopy (UPS) measurements were performed for separately prepared stacks of Si substrate (wet-oxidized)/X (5 nm) (including Co). The UPS measurements were performed using He I radiation ($h\nu$ = 21.2 eV) from a gas-discharge lamp. The base pressure of the chamber was $2 \times 10^{-8}$ Torr. Prior to the measurement, Ar ion sputtering was performed to remove any native oxides formed during the exposure to air. The metallic films were sputtered repeatedly until the Fermi edge was observed. For X = Cu (the lightest element investigated), the measurement was not satisfactory, owing to the significant damage induced to Cu during the Ar ion sputtering process. Therefore, a thicker layer (20 nm) was used in this case. More details on the measurement of the work function and the photoemission spectra are provided in Supplementary Note 2.

**Results and discussion**

In Fig. 3a–c, $K_1$ is plotted as functions of the work function ($W$), electronegativity ($\chi$), and spin–orbit coupling constant ($\xi$), respectively, all of which are taken from the literature[27–29]. We choose these material parameters because of their potential correlation with the inversion asymmetry or Rashba effect at the Co/X interface[30–32]. To estimate the strength of the correlation, we calculate Pearson's $r$ for all the plots. The Pearson's $r$ is close to ±1 (0) for a strong (weak) correlation. We obtain the correlation coefficients of 0.82, 0.63, and 0.07 for the plots in Fig. 3a, b, and c, respectively, indicating the strongest correlation between $K_1$ and $\Delta W$ ($\equiv W_X - W_{Co}$). $K_1$ also appears to be correlated with $\chi$ (Fig. 3b). This is expected, because the difference in $\chi$ between two elements is proportional to the charge transfer[33], which could be driven by the potential gradient at the Co/X interface in our samples. We note that this correlation feature is in accordance with a recent experimental observation for the interfacial Dzyaloshinskii–Moriya interaction originating from the inversion asymmetry[27]. We also plot $K_1$ as a function of $\Delta W$ measured for our samples by UPS (denoted as $\Delta W_{meas}$) (see Fig. 3d, Methods, and Supplementary Note 2 for details) and find a similar correlation between the two parameters ($K_1$ and $\Delta W_{meas}$), with a correlation coefficient of 0.81. This result shows that the inversion asymmetry at the interface plays an important role in the $K_1$ of Pt/Co/X heterostructures.

Figure 3e–h shows the results for $K_2$, which is similar to those for $K_1$ shown in Fig. 3a–d. The correlation coefficients for $K_2$ are –0.59, –0.51, and –0.18 for literature values of $\Delta W$, $\chi$, and $\xi$, respectively. Similar to $K_1$, $K_2$ exhibits meaningful correlations with $\Delta W$ and $\chi$. The correlation coefficient of $K_2$ with $\Delta W_{meas}$ is substantially improved to –0.94 (Fig. 3h), suggesting a strong correlation. Importantly, $K_2$ changes its sign depending on the type of material X but still shows an almost linear correlation with $\Delta W_{meas}$. According to this result, we conclude that the inversion asymmetry is an intrinsic origin of $K_2$ in Pt/Co/X heterostructures. We call it intrinsic because this mechanism is distinct from the first (spatial fluctuations of $K_1$[18]) and second (interfacial PMA combined with a gradual weakening of the exchange energy along the thickness direction[19])

mechanisms, which are extrinsic. Furthermore, our simple tight-binding model calculation with Rashba spin–orbit coupling supports this conclusion, as it shows that $K_2$ can have both positive and negative signs depending on the band filling even though $K_1$ is positive (i.e., PMA) (Supplementary Note 3).

The correlation result suggests that a large negative $\Delta W$ results in a large positive $K_2$, which is needed to form the easy-cone state. For experimental realization, we replace the metallic X layer with an MgO layer (see Supplementary Note 4). We choose MgO for the following two reasons. First, strong Rashba splitting was observed at metal–oxide interfaces[31,32]. Our $\Delta W_{\text{meas}}$ value at the Co/MgO interface is consistent with this expectation: it is –0.36 eV (Supplementary Note 4), which is more negative than the value (–0.25 eV) for the Co/Ta interface, which exhibits the most negative $\Delta W_{\text{meas}}$ among all the metallic Co/X interfaces. Second, MgO is widely adopted in various spintronic devices[3]. For a Pt/Co (1.0 nm)/MgO stack, we obtain $K_1$ of $1.47 \times 10^7$ erg/cm$^3$ and $K_2$ of $2.61 \times 10^6$ erg/cm$^3$. Compared with the all-metallic structures, the $K_2$ of the Pt/Co/MgO structure is larger by an order of magnitude, which is in accordance with our conclusion in this work; the inversion asymmetry is an intrinsic origin of $K_2$. However, previously proposed mechanisms[18–20] not considering the role of the inversion asymmetry are unable to explain the enhanced $K_2$ (see Supplementary Note 6 for details). Nonetheless, we note that the simple linear correlation between $K_2$ and $\Delta W_{\text{meas}}$ describes the enhanced $K_2$ of the Pt/Co/MgO structure only qualitatively, not quantitatively. Extrapolation of the linear line in Fig. 3h gives a $K_2$ value of approximately $0.27 \times 10^6$ erg/cm$^3$, which is significantly smaller than the measured value of $2.61 \times 10^6$ erg/cm$^3$. This large deviation may indicate that $\Delta W_{\text{meas}}$ is not the sole factor determining the inversion asymmetry for a metal–oxide interface. A recent experimental work combined with a first-principles study found that the asymmetric charge-density distribution (or the charge transfer) at a metal–oxide interface has an larger effect on the Rashba splitting than the work-function difference (or the potential gradient)[32].

This large and positive $K_2$ allows the easy-cone state to be formed in Pt/Co/MgO structures at $t_{\text{Co}}$ near the spin reorientation transition[1]. The formation of the easy-cone state is validated by both vibrating sample magnetometry and AHE measurements (Supplementary Note 7). The $K_1^{\text{eff}}$ and $K_2$ values for the Pt/Co (1.80–2.05 nm)/MgO structures are overlaid on a magnetic phase diagram (Fig. 4a). The cone angle ($\theta_c$) is estimated according to the relationship $\theta_c = \sin^{-1}(\sqrt{-K_1^{\text{eff}}/2K_2})$. We find that $\theta_c$ can be engineered over a wide range by controlling $t_{\text{Co}}$ (Fig. 4b), which is beneficial for device applications of the easy-cone state.

**Conclusion**

We investigated the origin of $K_2$ in Pt/Co/X heterostructures and found that the inversion asymmetry plays an important role in $K_2$. Among the material parameters considered in this study, the work-function difference at the Co/X interface shows the strongest correlation with both $K_1$ and $K_2$. Replacing the metallic X layer with MgO, whose interface with Co has a strong inversion asymmetry, we obtain greatly enhanced $K_2$, allowing the easy-cone state. The intrinsic origin of $K_2$ revealed in this study will contribute to the control of its values and therefore allow various easy-cone states suitable for a wide variety of spintronic applications.

**Acknowledgements**

This research was supported by the Creative Materials Discovery Program through the National Research Foundation of Korea (No. 2015M3D1A1070465), the Samsung Electronics University R&D program, and the KIST Institutional Program (Project Nos. 2V05750, 2E29410).

**Author contributions**

H.K.G. and S.H.L. planned and designed the experiment. H.K.G. prepared the samples and performed measurements. H.-J.P., K.-W.K., and K.-J.L. performed theoretical analysis. All authors discussed the results and contributed to the manuscript.

**Competing interests**

The authors declare no competing interests.

**Figure captions**

**Figure 1 | Phase diagram showing various magnetic states and inverse thickness dependences of $K_1$ and $K_2$.** **a**, Magnetic phase diagram as functions of $K_1^{\text{eff}}$ and $K_2$, showing four different magnetic states and their energy surfaces. **b, c,** Inverse Co thickness dependence of $K_1$ (**b**) and $K_2$ (**c**) for the Pt/Co ($t_{\text{Co}}$)/Cu structure. The error bars in $K_1$ and $K_2$ were obtained from three repeated measurements. The negative slope in **c** indicates negative interfacial $K_2$.

**Figure 2 | Measurement of magnetic anisotropy.** **a**, Schematic showing the Hall bar device used for the magnetic anisotropy measurements, together with an optical microscopy image (upper right). The Hall voltage was measured while injecting an in-plane current ($I_x$) along the $x$ direction. The $H_{\text{ext}}$ was applied along $\theta_H = 80°$ to facilitate a coherent magnetization behaviour. **b**, $m_z$–$H_{\text{ext}}$ plots for Pt/Co/X heterostructures. The symbols and the dashed lines indicate the results of AHE measurements and macrospin simluations, respectively. **c**, $\alpha H_{\text{ext}}$ vs. $1-m_z^2$ plots converted from the results in **b**. The solid lines represent the linear fittings to the data.

**Figure 3 | Correlation of magnetic anisotropies with material parameters. a–d,** $K_1$ as a function of $\Delta W$ (**a**), $\chi$ (**b**), $\xi$ (**c**), and $\Delta W_{\text{meas}}$ (**d**) for Pt/Co/X stacks with various X elements. **e–h,** Correlation results for $K_2$, similar to those shown in **a–d**. The values of $\Delta W$, $\chi$, and $\xi$ were taken from the literature[27–29], and those of $\Delta W_{\text{meas}}$ were obtained in this study via UPS measurements. The error bars of $K_1$, $K_2$, and $\Delta W_{\text{meas}}$ were obtained from three repeated measurements, whereas those of $\Delta W$ represent the standard deviations of reported values.

**Figure 4 | Pt/Co/MgO structure with easy-cone state. a,** Magnetic phase diagram in the $K_1^{\text{eff}}$–$K_2$ plane, with the $K_1^{\text{eff}}$ and $K_2$ values indicated at several $t_{\text{Co}}$ values in nanometers. **b,** Plot of $\theta_c$ vs. $t_{\text{Co}}$. The error bars of $K_1^{\text{eff}}$, $K_2$, and $\theta_c$ were obtained from three repeated measurements.

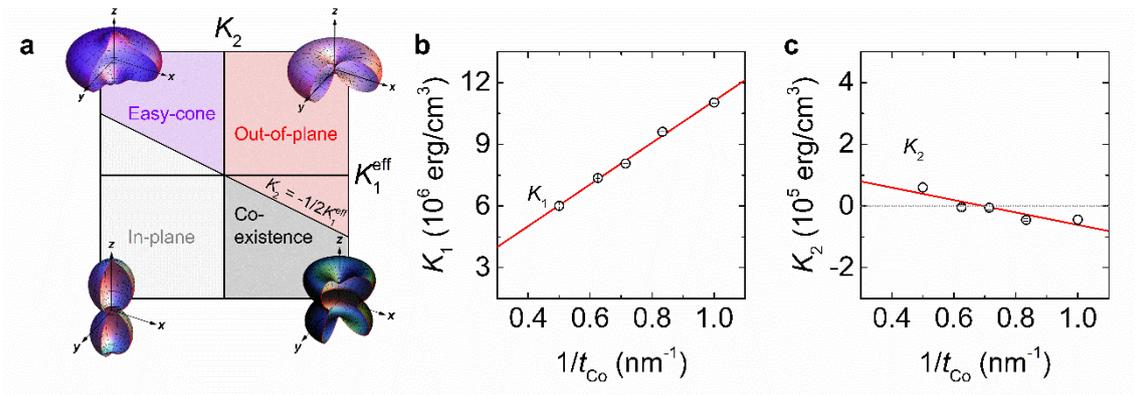

Figure 1.

Hyung Keun Gweon et al.

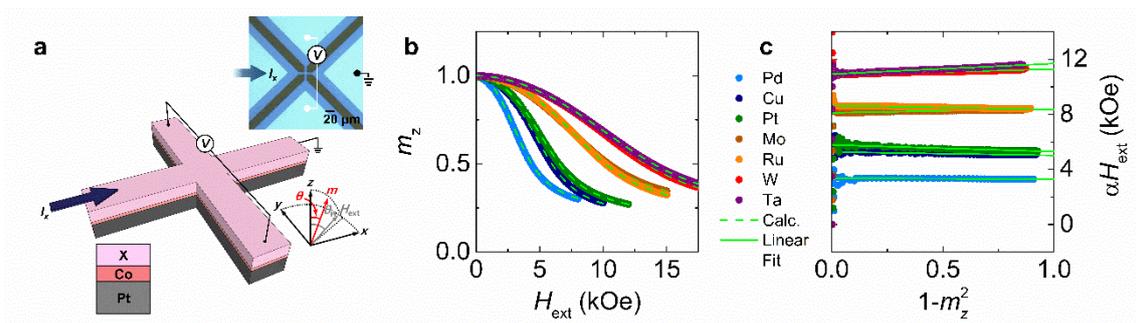

Figure 2.

Hyung Keun Gweon et al.

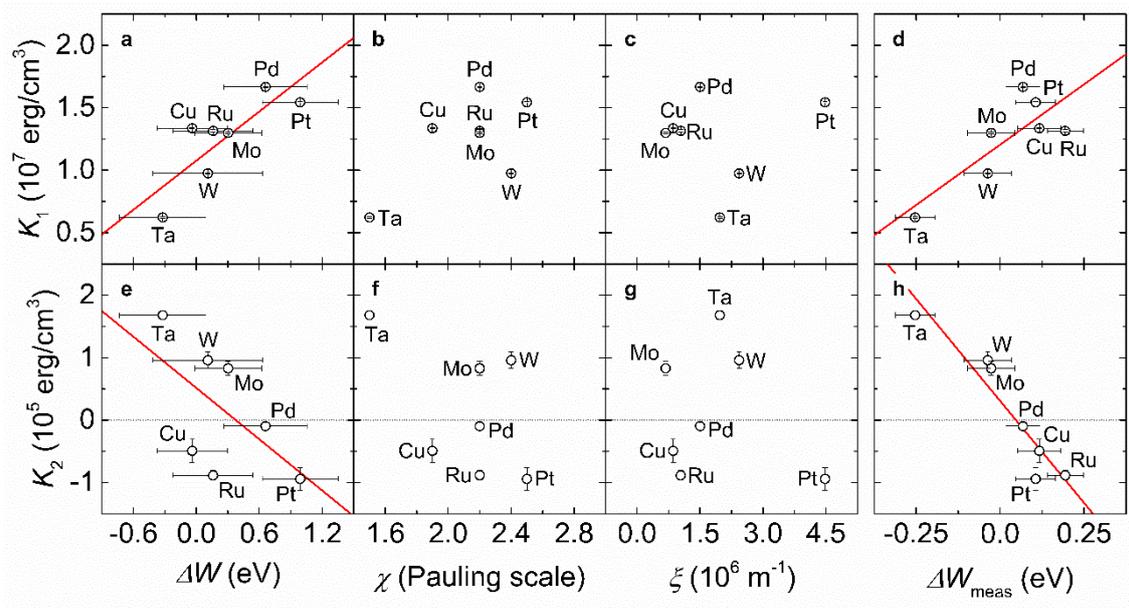

Figure 3.

Hyung Keun Gweon et al.

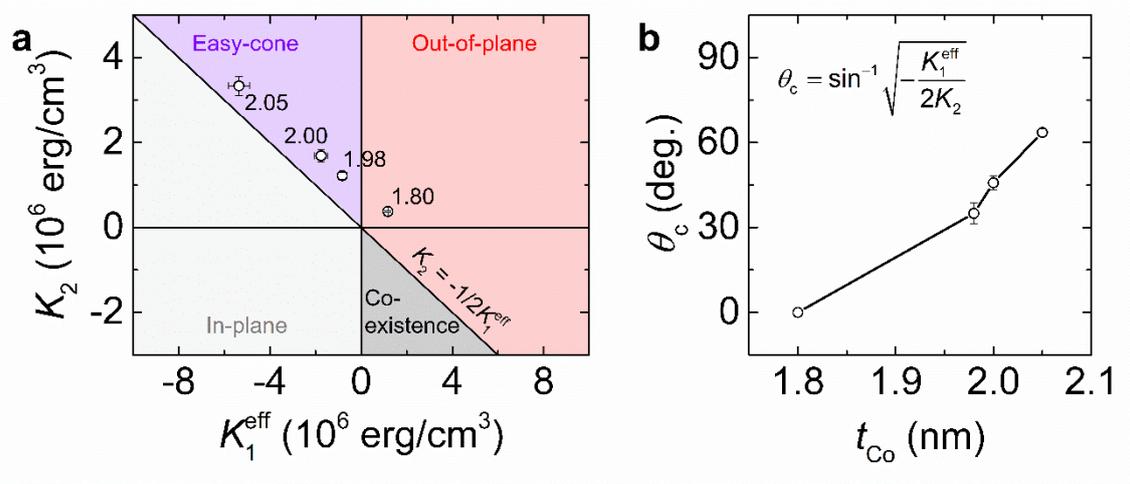

Figure 4.

Hyung Keun Gweon et al.